\documentclass[prl,amsmath,twocolumn,longbibliography]{revtex4-1}
\usepackage{amsmath}
\usepackage{amssymb}
\usepackage{amsthm}
\usepackage{graphicx}
\usepackage{stmaryrd}
\usepackage{siunitx}
\usepackage{ulem}
\usepackage[usenames,dvipsnames]{xcolor}

\usepackage{mathtools}
\usepackage{pgfplots}
\pgfplotsset{/pgf/number format/use comma,compat=newest}
\usepackage{algorithm}
\usepackage{algorithmic}
\usepackage[algo2e, linesnumbered, ruled, vlined]{algorithm2e}
\usepackage{mathtools}
\usepackage{enumitem}
\usepackage{booktabs}

\renewcommand\epsilon{\varepsilon}

\usepackage{physics}

\usepackage{tikz}

\newcommand{\obs}[1]{\ensuremath{\langle #1 \rangle}}
\newcommand{\pure}[1]{\ensuremath{| #1 \rangle \langle #1 |}}

\usepackage[hidelinks]{hyperref}
\begin{document}

%Title of paper
\title{Quantum Split Neural Network Learning using Cross-Channel Pooling}

\author{Won Joon Yun}
\affiliation{School of Electrical Engineering, Korea University, South Korea}
\author{Hankyul Baek}
\affiliation{School of Electrical Engineering, Korea University, South Korea}
\author{Joongheon Kim}
\affiliation{School of Electrical Engineering, Korea University, South Korea}

\date{\today}

\begin{abstract}
In recent years, the field of quantum science has attracted significant interest across various disciplines, including quantum machine learning, quantum communication, and quantum computing. Among these emerging areas, quantum federated learning (QFL) has gained particular attention due to the integration of quantum neural networks (QNNs) with traditional federated learning (FL) techniques.
In this study, a novel approach entitled quantum split learning (QSL) is presented, which represents an advanced extension of classical split learning. Previous research in classical computing has demonstrated numerous advantages of split learning, such as accelerated convergence, reduced communication costs, and enhanced privacy protection.
To maximize the potential of QSL, cross-channel pooling is introduced, a technique that capitalizes on the distinctive properties of quantum state tomography facilitated by QNNs. Through rigorous numerical analysis, evidence is provided that QSL not only achieves a 1.64\% higher top-1 accuracy compared to QFL but also demonstrates robust privacy preservation in the context of the MNIST classification task.
\end{abstract}

\maketitle
 \textit{Introduction.}
As quantum supremacy and recent advancements in quantum computing emerge, quantum algorithms are increasingly recognized as promising solutions for various computation-intensive problems that require rapid information processing~\cite{arute2019quantum,farhi2014quantum}. Notably, Shor's algorithm \cite{Shor97} enhances factorization computations, while Grover's search algorithm \cite{Grover96} reduces searching costs, both demonstrating quantum supremacy with real quantum devices. IBM Quantum's 2022 roadmap projects the development of 10,000 to 100,000 qubit quantum computers by 2026~\cite{roadmap2022}, further underscoring the potential for quantum algorithms to revolutionize next-generation computing and stimulate innovation in quantum machine learning (QML)~\cite{gao2018experimental,levine2019quantum,lu2020quantum, huang2021information, liu2018quantum}.

Recent research has begun to reimplement existing ML applications using quantum neural networks (QNNs), such as image classification~\cite{baek2022scalable,dilip2022data}, reinforcement learning~\cite{yun2022quantum,icdcs2022yun,wauters2020reinforcement}, and federated learning (FL)~\cite{yun2022slimmable,KWAK2022}. In particular, quantum federated learning (QFL) has gained traction in QML, as it enables distributed learning through quantum communications\cite{xu2022privacy,li2021quantum}. Inspired by these developments and the success of split learning (SL) in privacy preservation~\cite{DSN21}, the classical SL framework is extended to the quantum SL (QSL) domain.

A na\"{i}ve approach to QSL might simply reimplement the classical SL using quantum principles. However, this approach neglects the unique properties of quantum phenomena. As reported in \cite{baek2022scalable,baek2022sqcnn3d,baek2022fv}, features obtained from quantum convolution neural networks (QCNNs) via naïve training exhibit similarities when inputs are provided to the QCNNs. To address this issue, reverse fidelity training has been proposed to leverage quantum properties effectively. In designing QSL, such quantum characteristics can contribute valuable insights to QML.

The primary objective is to propose an extension of SL to QSL by substituting neural networks with QNNs~\cite{DSN21}. Furthermore, cross-channel pooling (C2Pool) is introduced, which capitalizes on the expected value of probability amplitude inherent in QNN outputs. Empirical analysis demonstrates that incorporating C2Pool results in superior accuracy, minimized communication costs, and robust privacy preservation. The contributions of this paper are threefold: (1) The quantum split learning framework is presented for the first time; (2) The proposed framework considers the unique characteristics of quantum phenomena; and (3) Numerical experiments reveal that the framework outperforms both na\"{i}ve QSL and QFL approaches.

\textit{Quantum Neural Networks.}
Before delving into the concept of quantum neural networks, it is essential to first introduce the fundamental quantum operations. The basic quantum gates will be discussed in the context of a single-qubit system. In such a system, the quantum state is defined using two bases, $|0\rangle=[1~0]^T, |1\rangle=[0~1]^T$, and the probability amplitude within the Hilbert space $\mathcal{H}$ as $
|\psi\rangle = \alpha_0 |0 \rangle + \alpha_1 |1 \rangle$, where $\sum\nolimits^1_{i=0}|\alpha_i|^2 = 1$. The quantum state for a $Q$-qubit system is expressed as $
\ket{\psi} = \sum^{2^Q-1}{n=0} \alpha_n \ket{n}$,
where $\alpha_n$ and $\ket{n}$ represent the probability amplitude and the $n$-th standard basis in Hilbert space, respectively. Owing to the definition of Hilbert space (\textit{i.e.}, $\mathcal{H}^{\otimes Q} \equiv \mathbb{C}^{2^Q}$), the probability amplitude $\forall \alpha_n \in \mathbb{C}$ maintains the following relationship: $\sum^{2^Q-1}{n=0} |\alpha_n|^2 = 1$.

The \textit{rotation} and \textit{controlled} gates linearly transform the probability amplitude and are denoted as
\begin{align}
&R_\Gamma (\theta) \!=\! \exp(\!\!-i \frac{\theta}{2} \Gamma\!\!),
C({\Gamma})\!=\!
\begin{bmatrix}I & 0 \\ 0 & \Gamma\end{bmatrix}, \forall \Gamma \!\in\! \{X,Y,Z\},\label{eq:gate-basic}\\
&\text{s.t.~}I\!=\!\begin{bmatrix} 1 &  0 \\ 0 &  1\end{bmatrix},
X\!=\!\begin{bmatrix} 0 &  1 \\ 1 &  0\end{bmatrix}, Y\!=\!\begin{bmatrix} 0 & -i \\ i &  0\end{bmatrix}, 
Z\!=\!\begin{bmatrix} 1 &  0 \\ 0 & -1\end{bmatrix},\nonumber%\label{eq:gate-pauli}
\end{align}
where $X$, $Y$, and $Z$ denote the Pauli-$X$, $Y$, and $Z$ gates, respectively.

\begin{figure}[t!]
\centering
\includegraphics[width=\columnwidth]{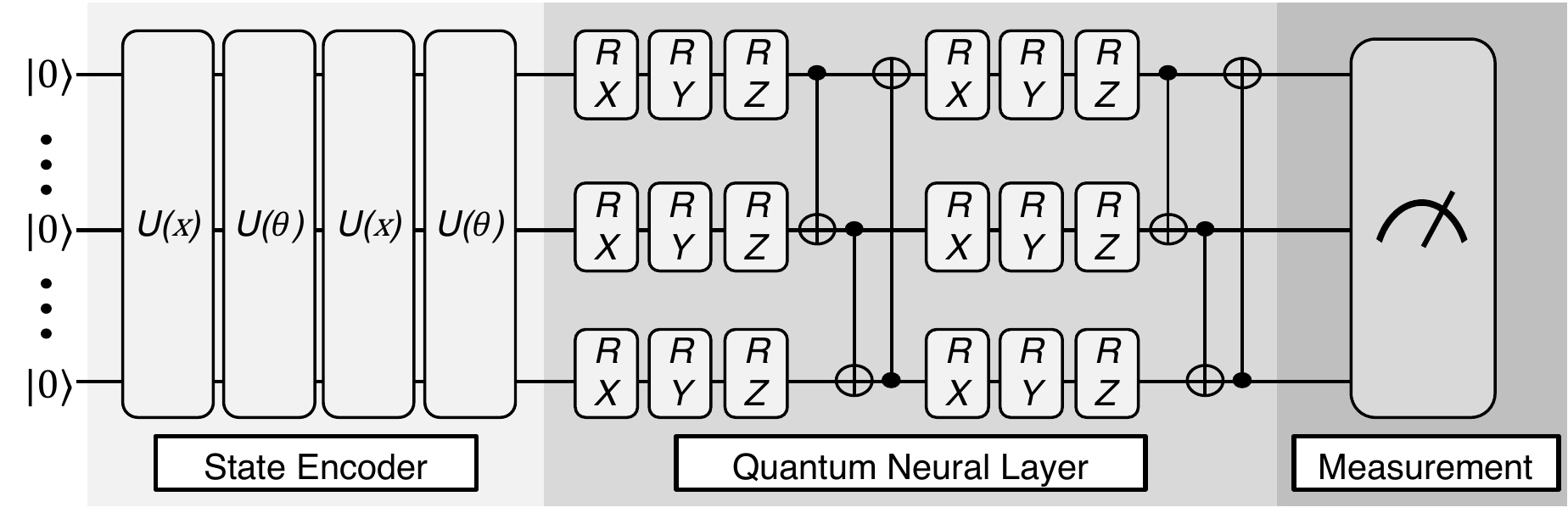}
\caption{The general quantum neural network architecture.}
\label{fig:QNN}
% \vspace{-10pt}
\end{figure}

 As illustrated in Fig.\ref{fig:QNN}, a QNN is divided into three parts: the state encoder, quantum neural layer, and measurement~\cite{qoc2022wang}.
 Uploading classical data to quantum circuits is a challenging task, prompting the exploration of various encoding methods for QNNs, such as basis encoding, amplitude encoding, and angle encoding. Among these methods, an advanced angle encoding technique, known as data-reuploading, is introduced. Given the classical data $\mathbf{x} \in \mathbb{R}^{|\mathbf{x}|}$, the data-uploading process is expressed as:
\begin{equation}
    |\psi_{\textit{enc}}\rangle = \Big(\prod\nolimits_{l=0}^{ \lceil |\mathbf{x}| / Q\rceil }U(\theta_l)U(\mathbf{x}_{Q \cdot l:Q\cdot(l+1)})\Big)\cdot |\psi_0\rangle \label{eq:data_reuploading}. 
\end{equation} Here, $|\psi_0\rangle$ denotes the initial quantum state, which is $|\psi_0\rangle = |0\rangle^{\otimes Q}$. The input data $\mathbf{x}$ are divided into segments $\mathbf{x}_{Q\cdot l:Q\cdot (l+1)}$ with an interval of $Q$, where $\mathbf{x}_{Q\cdot l:Q\cdot (l+1)}$ and $\theta_l$ represent the input vector composed of the first $Q \cdot l$ to $Q \cdot(l+1)$ elements and the trainable parameters, respectively.
Quantum neural layers are responsible for processing the quantum state $|\psi_{\textit{enc}}\rangle$ and can be employed as a quantum convolutional neural network (QCNN) or a quantum fully connected network (QFCN). The rotation gates and controlled gates, as defined in \eqref{eq:gate-basic}, are used to construct the quantum neural layers. Designing a quantum neural layer is crucial for real quantum devices, as entangling $Q$ qubits is challenging. However, classical computers can simulate quantum circuits, enabling the quantum neural layer to be regarded as a unitary operation, \textit{\textit{i.e.}}, $|\psi\rangle = U(\theta) |\psi_{\textit{enc}}\rangle$.
To obtain the classical output, the quantum state must be projected onto a measurement device. Various measurement methods are introduced, such as projection-valued measure (PVM), positive operator-valued measure (POVM), and trainable measurement method. PVM projects the quantum state into a projector ${\pure{n}}_{n=1}^{2^q}$, producing a probability measure with the following output:
    \begin{equation}
    \Pr(y=n) = \langle \psi| n \rangle  \langle n |\psi \rangle =  |\langle \psi| n \rangle |^2 = |\alpha_n|^2. \label{eq:pvm}
    \end{equation}
 PVM is considered a \textit{complete} measurement since it projects the quantum state into all possible bases. Typically, the quantum state is projected into Pauli-$Z$ measure devices for each qubit, where Pauli-$Z$ is defined as ${Z} = \text{diag}(1, -1)$. The $n$-th projection matrix is defined as ${P}_n \triangleq {I}^{\otimes n-1}\otimes {Z} \otimes {I}^{\otimes Q-n}$. In POVM, the expectation value of a projection ${P}_n$ is denoted as a random variable $O_n \in {R}[-1, 1]$. The expectation value of the $m$-th qubit is given by $\obs{O_n} = \bra{\psi} {P}_n \ket{\psi}$. POVM is known as an \textit{incomplete} measurement, as it projects the quantum state into partial projection matrices.
Trainable measurement, as suggested by Lloyd et al. (2020) and Schuld et al. (2022), is another method to consider. To implement trainable measurement, the concept of a pole is proposed, which involves the use of multi-agent reinforcement learning and federated learning. This approach allows for greater adaptability and potential improvements in measurement outcomes, making it a valuable addition to the quantum computing toolbox.

\textit{Optimizing Methods for QNN.}
A quantum neural network (QNN) training method known as the classical-quantum hybrid method is introduced. This approach is necessary due to the \textit{backward quantum propagation of phase errors (Baqprop) principle}, which requires classical computing to obtain the gradient of the objective function. The classical-quantum hybrid method consists of two main processes: (1) gradient calculation using classical computing, and (2) the parameter-shift rule.

Let the loss function be denoted as $\mathcal{L}(\Theta)$, where $\Theta$ represents the QNN parameters. The loss gradient of the $i$-th QNN parameter can be expressed as $\frac{\partial \mathcal{L}(\Theta)}{\partial \theta_{i}} = \frac{\partial \mathcal{L}(\Theta)}{\partial f(\Theta)} \cdot \frac{\partial f(\Theta)}{\partial \theta_{i}}$, where $f(\Theta)$ denotes the measured output. The classical computer calculates the term $\frac{\partial \mathcal{L}(\Theta)}{\partial f(\Theta)}$, while the term $\frac{\partial f(\Theta)}{\partial \theta_{i}}$ is computed by quantum computers using forward propagation as,
\begin{equation}
\frac{\partial f(\Theta)}{\partial \theta_{i}}=  f\Big(\Theta + \frac{\pi}{2} \mathbf{e}_{i}\Big) - f\Big(\Theta - \frac{\pi}{2} \mathbf{e}_{i}\Big), \label{eq:ps-rule}
\end{equation}
 where $\mathbf{e}_{i}$ is one-hot vector that eliminate all components except $\theta_{i}$, \textit{i.e.}, $\Theta \cdot \mathbf{e}_{i} = \theta_{i}$.  

\textit{Recent Studies on Quantum Distributed Learning.}
Recent studies in quantum distributed learning, including quantum federated learning, have been introduced. QFL is an architecture inspired by classical federated learning and enhanced by incorporating quantum machine learning. The QFL model shares many similarities with classical FL, as it inherits numerous characteristics from the classical version. In QFL, multiple devices and a server participate in transmitting aggregated model parameters repeatedly.
Additionally, QFL with quantum blind computing has been proposed to strengthen privacy-preserving features by employing differential privacy in conjunction with quantum computers. However, due to qubit constraints, the application of QFL is significantly limited. In \cite{chen2021federated,huang2022quantum}, binary classification is utilized in the simulation environment, representing a relatively simple task. This limitation exists because exploiting spatial information demands a substantial number of qubits. To address this issue, hybrid-QFL is proposed in \cite{chen2021federated}, where a pretrained feature extractor is employed to extract low-dimensional features from images.
Conversely, the current work concentrates on overcoming these QFL challenges, which will be discussed further in next.

\begin{figure*}[t!]
\centering
\includegraphics[width=\textwidth]{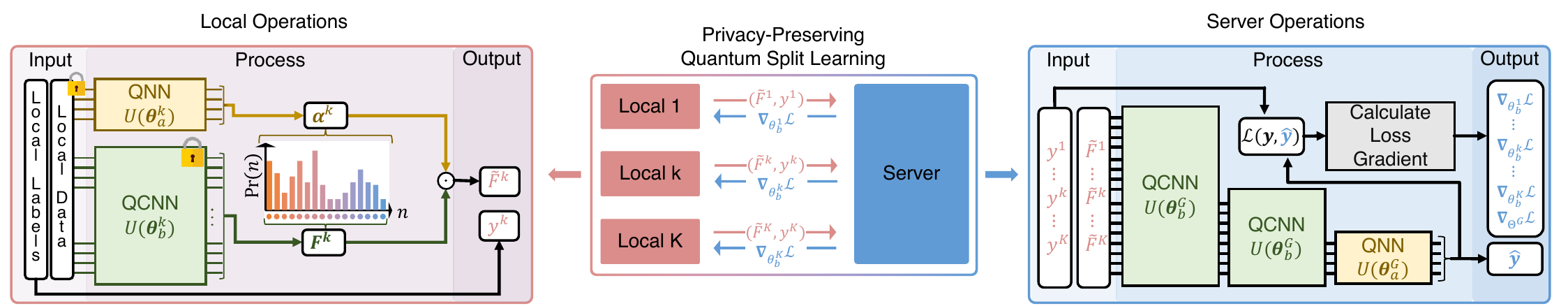}
\caption{The illustration of privacy-preserving quantum split learning framework.}
\label{fig:main}
% \vspace{-8pt}
\end{figure*}
\textit{Quantum Split Learning.}
The setting of QSL is as following. There are $K$ local devices and one server participating in the proposed quantum split learning. As illustrated in Fig.~\ref{fig:main}, each local device $k$ has its local data $\mathbf{Z}_k \in \mathbf{Z}$ and quantum neural networks. Every local device possesses $L$ QCNNs and one cross-channel pooling QNN. The local data and local QCNNs cannot be shared with others, but features and labels can be. The server processes the features and eventually generates corresponding predictions.

Next, the local-side operations are elaborated upon. Initially, quantum image processing is carried out. The classical data of local $k$ is represented as $\mathbf{X}^k \in \mathbb{R}^{W \times H \times c_{\text{in}}}$, and the QCNN filters scan the image with a stride of $s$, where the filter size is $\kappa$. In this paper, the patch $\mathbf{x}^k \in \mathbb{R}^{\kappa \times \kappa \times c_{\text{in}}}$ is described, with $c_{\text{in}}$ denoting the number of output channels. The patch $\mathbf{x}^k$ is uploaded to QCNN using the data-reuploading method referenced in \eqref{eq:data_reuploading}. A scalable quantum convolutional neural network (sQCNN \cite{baek2022scalable}) is adopted as the QCNN. By utilizing the QCNN and a patch $\mathbf{x}^k$, the feature $\mathbf{f}^k \in \mathbb{R}^{c_{\text{out}}}$ is obtained, where $c_{\text{out}}$ indicates the number of output channels.

Subsequently, a novel cross-channel pooling of QNN is proposed. In a similar manner, the probability amplitude is acquired through QNN, and the output is denoted as $ \alpha^k(\mathbf x^k) \in \mathbb{C}^{{c_{\text{out}}}}$, where $|{\alpha}^k (\mathbf x ^ k)|^2 = 1$. A cross-channel pooling method is proposed by taking the inner product of $\mathbf{f}^k$ and $\alpha^k(\mathbf x^k)$. Consequently, the output is expressed as follows:
\begin{equation}
    \tilde{f}^k = \mathbf{f}^k \cdot \alpha^k(\mathbf x^k). \label{eq:feature}
\end{equation}
After scanning all patches, the output features are obtained as $\tilde{\mathbf{F}}^k = { \tilde{f}^k } \in \mathbb{R}^{\hat W \times \hat H \times 1}$. Then, the local device $k$ transmits its feature-label set $(\tilde{\mathbf{F}}^k, y^k)$ to the server. It is important to note that the feature-label set is a classical data, even though it is processed with QNN.

\begin{algorithm2e}[t]
\small
    % \SetCustomAlgoRuledWidth{0.44\textwidth}  
\caption{Quantum Split Learning}
\label{alg:qsl} 
~\textbf{Initialize.} $\mathcal{Z}\leftarrow \emptyset$, $\forall \Theta^k$, and $\Theta^G$\;
~\textbf{\texttt{$\triangleleft$ Local-side operations $\triangleright$}}\\
~\textbf{For} {$k \in \{1,\cdots, K\}$} \textbf{do} \\
{   ~~~~Generate features with local data ${\mathbf{X}}^k$\\
    ~~~~Obtain cross-channel pooled features $\tilde{\mathbf{F}}^k$ with \eqref{eq:feature}\\
    ~~~~Transmit features and labels to server\\
 } 
~\textbf{end} \\
~~\textbf{\texttt{$\triangleleft$ Server-side operations $\triangleright$}}\\
~\textbf{For} {$k \in \{1,\cdots, K\}$} \textbf{do}\\
{   
    ~~~~Make prediction with features and calculate loss\\
    ~~~~Calculate the local gradient, $\nabla_{\Theta^{k}} \mathcal{L}$\\
    ~~~~Distribute gradients to local $k$\\
 }
 ~\textbf{end} \\
 ~Calculate the global gradient, $\nabla_{\Theta^{G}} \mathcal{L}$\\
~\textbf{\texttt{$\triangleleft$ Parameters update $\triangleright$}}\\
~\textbf{Local $k$:~} $\Theta^k \leftarrow \Theta^k - \eta \nabla_{\Theta^{k}} \mathcal{L}$\;
~\textbf{Server:} $\Theta^G \leftarrow \Theta^G - \frac{\eta}{K} \nabla_{\Theta^{G}} \mathcal{L}$\;
\end{algorithm2e}

\textit{Server-Side Operations in Quantum Split Learning}.
This paper presents the server-side operations within the proposed quantum split learning framework. The server processes the features using QCNNs and QNN, with QCNN operations being identical to those on the local side. The QNN, on the other hand, is responsible for multi-class classification, employing the projection-valued measure (PVM), a quantum state projection method~\cite{yun2022projection}.

The prediction value is obtained when the quantum state $\ket{\psi}_{\mathbf{X},\Theta^G}$ is projected onto basis states, \textit{i.e.}, $\{\ket{0}, \cdots, \ket{n}, \cdots,\ket{2^q-1}\}$. Notably, this method does not rely on the softmax function, thus avoiding the need for softmax tuning parameters~\cite{jerbi2021variational}.

Next, the process of training the QNN is detailed, with the overall procedure outlined in Algorithm~\ref{alg:qsl}. The $k$-th local device initially generates cross-channel pooled features $\tilde{\mathbf{F}}_k$ from its data (lines 4--5) and transmits them alongside labels to the server (line 6). Upon receiving all features and labels, the server generates predictions via feed-forwarding.
For simplicity, the trainable parameters are denoted as $\Theta = \{\{\Theta^k\}_{k=1}^K,\Theta^G\}$. The objective function for PVM, as suggested by \cite{yun2022projection}, is defined as:
\begin{equation}
\mathcal{L}(\Theta; \zeta) =  \frac{1}{|\zeta|}\sum\nolimits_{(\mathbf{X}, \mathbf {y}) \in {\zeta}} [\mathcal{L}_{\text{BCE}}(\Theta;\mathbf{X}) + \mathcal{L}_{\text{PAR}}(\Theta;\mathbf{X})],
\end{equation}
where $\zeta$, $\mathcal{L}_{\text{BCE}}$, and $\mathcal{L}_{\text{PAR}}$ represent the minibatch, binary cross entropy, and regularizer term, respectively. The terms $\mathcal{L}_{\text{BCE}}$ and $\mathcal{L}_{\text{PAR}}$ are expressed as follows:
\begin{eqnarray} \label{eq:BCE}
\mathcal{L}_{\text{BCE}}(\Theta;\!\mathbf{X}) \! \! &=& \! \! -\! \!\! \sum^{|\mathbf{y}|}_{n=1}[y_n \!\log p_n \!\!+\!\! (1\!-\!y_n) \log (1\!-\!p_n)], \\
\mathcal{L}_{\text{PAR}}(\Theta;\!\mathbf{X})\! \!  &=&\! \!  -\! \!\!\sum^{2^q}_{n'>|\mathbf{y}|}\log (1\!-\!p_{n'}), \label{eq:PAR}
\end{eqnarray}
where $p_n$ denotes the prediction value for class $n$ obtained by \eqref{eq:pvm}. Lastly, the loss gradients of local QNNs and the global QNN are calculated and updated using \eqref{eq:ps-rule} (lines 10--12).

\textit{Evaluation Setup.}
The evaluation setup for the proposed quantum split learning framework comprises several benchmark schemes. First, the advantage of QSL is assessed by comparing the proposed framework to QFL with federated averaging, referred to as `QFedAvg', and standalone training. Top-1 accuracy is adopted as the evaluation metric, and for QFedAvg, 10 local iterations per communication round are employed.
Second, the effectiveness of cross-channel pooling is examined through an ablation study. The comparison frameworks include QSL without any pooling and QSL with channel-averaging pooling. Top-1 accuracy and communication cost are the metrics used to measure the performance of these approaches.
Lastly, scalability is evaluated since the number of local devices plays a crucial role in distributed machine learning. Various numbers of local devices are considered, specifically $K\in \{1,2,5,10,20\}$.
Experiments are conducted on 10-class classification tasks using MNIST, FashionMNIST, and CIFAR10 datasets, with all data bicubic interpolated to a size of $14 \times 14$. It is important to note that all data are independent and identically distributed. Each local device is allocated 512 data samples, which is also the batch size. Both local and server models consist of 312 trainable parameters, utilizing U3 and CU3 gates. The Adagrad optimizer with an initial learning rate of 1 is employed. All experiments are conducted on a classical computer using the Torchquantum and Python library.

\begin{table}[t]
\centering
\caption{Top-1 accuracy of various quantum distributed learning framework and dataset ($K=10$).}
\begin{tabular}{l|ccc}
    \toprule[1pt]   
      \textbf{Metric} & MNIST & FashionMNIST & CIFAR10\\\hline
    Standalone              & $58.13$   & $64.53$ & $14.68$ \\
    QFL                 & $63.32$   & $68.92$ & $22.89$ \\
    \textbf{QSL}  & $\mathbf{64.96}$   & $\mathbf{71.29}$ & $\mathbf{26.43}$ \\ \bottomrule[1pt]
\end{tabular}\label{tab:accuracy}
\end{table}
\begin{figure}[t!]\centering
\includegraphics[width=.8\columnwidth]{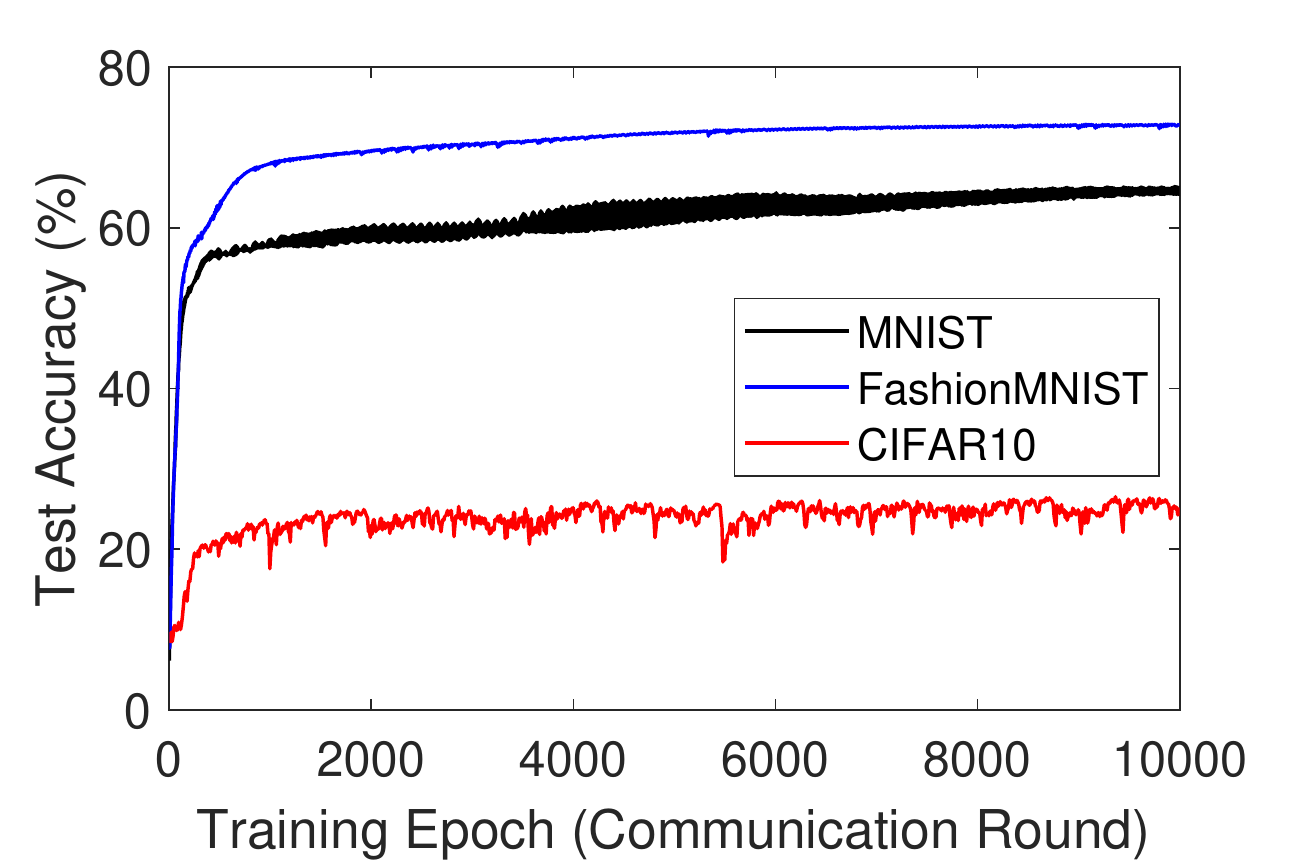}
\caption{Learning curve with various datasets.}
\end{figure}

\textit{Experimental Results.}
Table~\ref{tab:accuracy} shows the top-1 accuracy. Regarding MNIST dataset, quantum split learning shows 
1.64\% and 6.83\% higher than QFedAvg and standalone training. In addition, QSL shows 2.37\% and 6.76\% higher accuracy than QFedAvg and standalone training. Lastly, a similar tendency has been observed regarding CIFAR10. 
Thus, the proposed algorithm outperforms other algorithms.

Table~\ref{tab:ablation} shows the various results of C2Pool ablation. 
Concerning top-1 accuracy, QSL incorporating average-pooling and without pooling demonstrate inferior performance compared to QSL with C2Pool. Furthermore, QSL with C2Pool exhibits the smallest domain gap between training and test data. As for communication, the communication cost per feature is calculated as follows:
\begin{equation}
    \textit{Commun. Cost} = 4 \,\text{[Byte/float]} \cdot \hat{W} \times \hat{H} \times c_{\textit{out}} \, \text{[float]}. 
\end{equation}
The average pooling and C2Pool reduce 16 channels to a single channel, whereas QSL without pooling transmits the original feature that consists of 16 channels to the server, respectively. In summary, the proposed C2Pool improves performance as well as reduces gap and communication costs.

\begin{figure}[t!]
\centering
\includegraphics[width=.8\columnwidth]{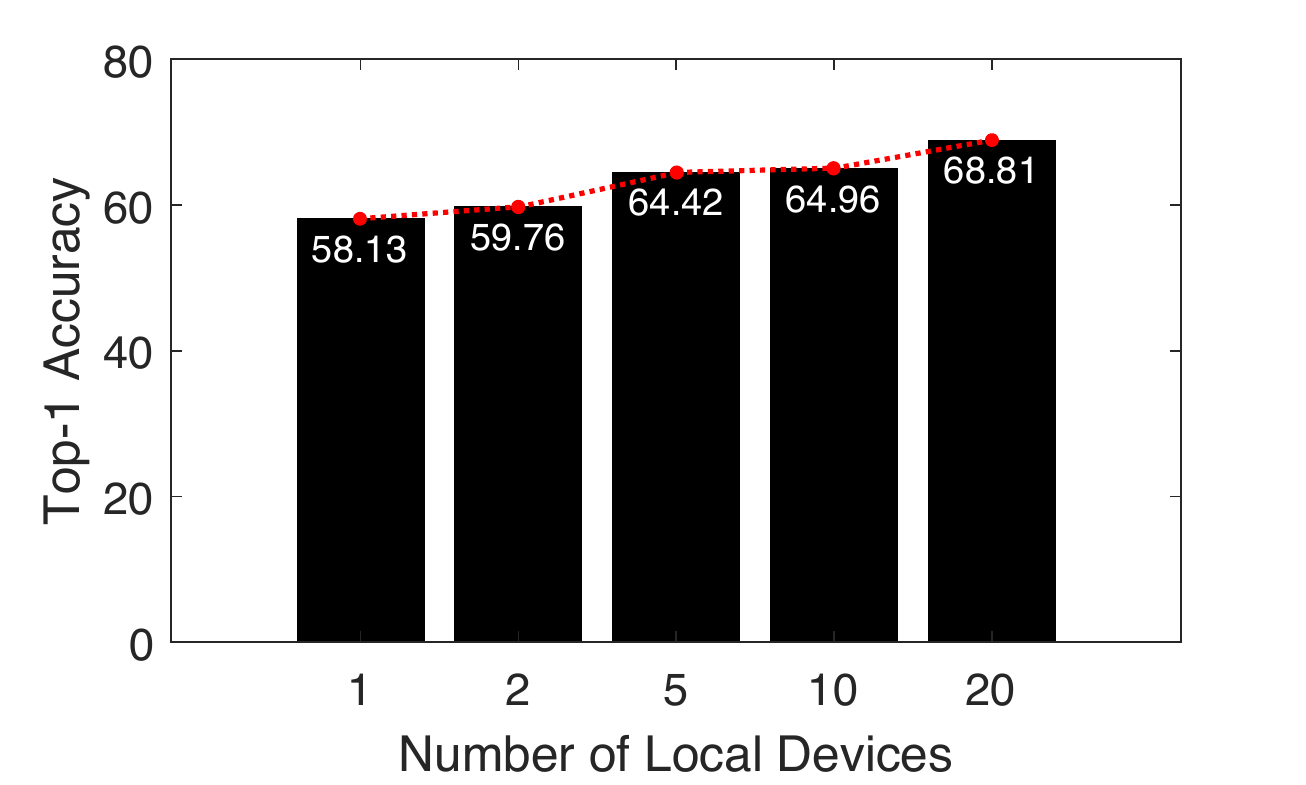}
\caption{Impact on the number of devices given MNIST dataset.}
\label{fig:scalability}
\end{figure}

Fig.~\ref{fig:scalability} shows the experiment that increase the number of devices from $K=1$ to $K=20$. When the number of local devices is 1, it means standalone training. When $K=20$, the top-1 accuracy achieves 68.81\%, which is the highest. Therefore, it can be confirmed that higher accuracy can be obtained, as the number of local devices increases. 

\begin{figure}[t!]
\centering
\includegraphics[width=\columnwidth]{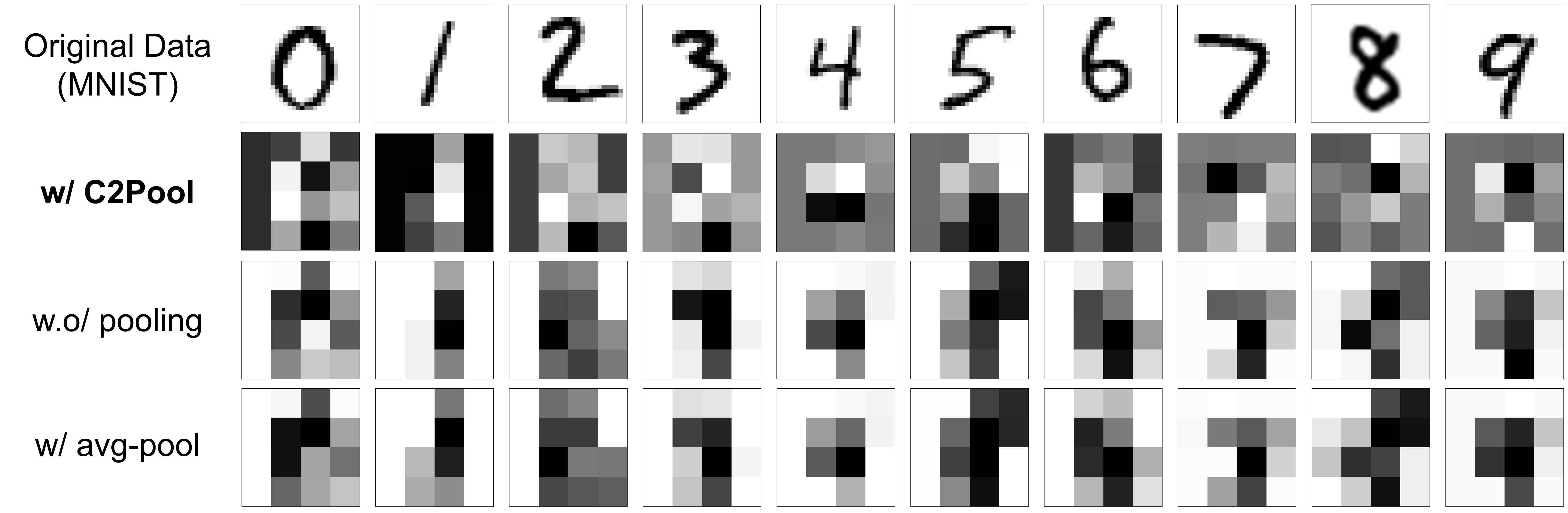}
\caption{The various output features that local client transmits to the server given MNIST image corresponding to different quantum split learning architectures.}
\label{fig:abstract}
% \vspace{-3mm}
\end{figure}

Fig.~\ref{fig:abstract} presents the various feature representation corresponding to different quantum split learning frameworks.  It is apparent that the all features
is harder recognizable than the original image, because the original image becomes distorted after passing through convolution. Especially in C2Pool, the outline of digits does not seem clear, whereas convolution without pooling and convolution with average-pooling clearly show 
 the outline. This is, the proposed C2Pool has a strength to privacy-preserving aspect.

\begin{table}[t!]
\centering
\caption{Ablation study on cross-channel pooling in the quantum split learning framework given MNIST dataset.}
% \scriptsize
\begin{tabular}{l|ccc}
    \toprule[1pt]   
      \textbf{Metric} & Train Acc. & Test Acc. & Commun. Cost [Byte]\\\hline
     w.o/ pooling              & $66.64$   & $59.45$ & $1,024$\\ 
     w/ avg-pooling            & $67.01$   & $58.32$  & $\mathbf{64}$ \\
    \textbf{w/ C2Pool}  & $\mathbf{68.96}$   & $\mathbf{64.96}$ & $\mathbf{64}$ \\ \bottomrule[1pt]
\end{tabular}
\label{tab:ablation}
\end{table}

\textit{Conclusions and Future Work.}
In this study, a quantum split learning framework was proposed to fully exploit the capabilities of quantum computing. A probability amplitude-based cross-channel pooling method was introduced to enhance the performance of the framework. Experiments demonstrated that the proposed framework achieves reasonable performance, scalability with respect to the number of local devices, low communication cost, and privacy preservation in 10-class classification tasks.
As this research represents the first quantum split learning framework, one potential direction for future work is the exploration of more detailed split learning approaches. Additionally, future work could involve investigating the differential privacy aspects of the proposed cross-channel pooling method.

\textit{Acknowledgments.}
This research was funded by the National Research Foundation of Korea (2022R1A2C2004869).
To whom correspondence should be addressed:  Joongheon Kim (e-mail: joongheon@korea.ac.kr)

\bibliography{reference}
\end{document}